\documentclass[aps,pre,superscriptaddress,twocolumn,balancelastpage]{revtex4-1}

\usepackage[colorlinks,bookmarks=false,citecolor=blue,linkcolor=blue,urlcolor=blue]{hyperref}
\usepackage[all]{hypcap}   

\usepackage{amsmath,amssymb}
\usepackage{graphicx}
\graphicspath{{figures/}{/}}

\usepackage{verbatim}
\usepackage{color}

\usepackage{placeins}  
\usepackage{flafter}     
\usepackage{float}

\usepackage{color}

\newcommand{\ue}{\text{e}}
\newcommand{\ui}{\text{i}}
\newcommand{\ud}{\text{d}}

\newcommand{\tb}{t_{\text{b}}}
\newcommand{\hbars}{\hbar_{\text{s}}}
\newcommand{\Dq}{D_{\text{q}}}
\newcommand{\HS}{H_{\text{S}}}
\newcommand{\HE}{H_{\text{E}}}
\newcommand{\rhoS}{\rho_{\text{S}}}
\newcommand{\rhoE}{\rho_{\text{E}}}

\makeatletter
\let\Hy@backout\@gobble
\makeatother

\begin{document}

\title{Linear and logarithmic entanglement production in an interacting chaotic system}

\author{Sanku Paul}
\affiliation{Max-Planck-Institut f\"ur Physik komplexer Systeme,
	N\"othnitzer Stra\ss e 38, 01187 Dresden, Germany}

\author{Arnd B\"acker}
\affiliation{Technische Universit\"at Dresden,
	Institut f\"ur Theoretische Physik and Center for Dynamics,
	01062 Dresden, Germany}
\affiliation{Max-Planck-Institut f\"ur Physik komplexer Systeme,
	N\"othnitzer Stra\ss e 38, 01187 Dresden, Germany}

\date{\today}

\begin{abstract}
    We investigate entanglement growth for a
    pair of coupled kicked rotors. For weak coupling, the growth of the
    entanglement entropy is found to be initially linear followed by a
    logarithmic growth. We calculate analytically the time after which the
    entanglement entropy changes its profile, and a good agreement
    with the numerical result is found.
    We further show that the different regimes of
    entanglement growth are associated with different rates of energy growth
    displayed by a rotor. At a large time, energy grows diffusively, which is
    preceded by an intermediate dynamical localization. The time-span of
    intermediate dynamical localization decreases with increasing coupling
    strength. We argue that the observed diffusive energy growth is the result
    of one rotor acting as an environment to the other which destroys the coherence.
    We show that the decay of the coherence is initially exponential followed 
    by a power-law.
\end{abstract}

\maketitle


Entanglement, as characterized by the von Neumann entropy,
has recently emerged as an indispensable tool to
distinguish phases and phase transitions in many-body quantum systems
and reveals highly non-local information
\cite{HorHorHorHor2009b,Laf2016,YanChaHamMuc2015}.
Many-body localization, which emerges due to ergodicity breaking, is known to
exhibit a logarithmic growth of the entanglement entropy
~\cite{BarPolMoo2012,SerPapAba2013,VosAlt2013,ZniProPre2008,LukRisSchTaiKauChoKheLeoGre2019,LuiLafAle2016}.
On the other hand, systems showing thermalization show a
saturation of entanglement growth \cite{SinBarPol2016}.
Logarithmic slow down is
also observed in a system with long-range interaction along with a quench
\cite{SchLanRooDal2013,LerPap2020,LerMarGamSil2019}
and also in many-body system with non-ergodic dynamics
arising due to glassy behavior \cite{BarPolMoo2012,vanLevGar2015}.

On the other hand, systems with interactions, for which a finite speed of
correlation spreading is generally observed~\cite{NahRuhVijHaa2017,SchLanRooDal2013,CalCar2005,LuiLafAle2016},
show a linear growth of the entanglement entropy
before saturating. Linear growth is
also observed in open quantum systems \cite{BiaHacYok2018,ZurPaz1994,MilSar1999},
for instance, an inverted harmonic
oscillator weakly coupled to a thermal bath \cite{ZurPaz1994}.
It is further conjectured in Ref. \cite{ZurPaz1994},
that the rate of linear growth equals the sum of positive Lyapunov
exponents of the system. Similar correspondence between entropy
production and the Lyapunov exponent
has been shown
for a kicked rotor coupled to a thermal bath \cite{MilSar1999}.
However, coupled kicked tops show a violation of this conjecture
and it is observed that the rate depends on the coupling strength
rather than on the Lyapunov exponents \cite{FujTanMiy2003, FujMiyTan2003}.

In addition to  many-body and open quantum systems,
even isolated two-body systems
are capable of showing a non-trivial and often unexpected dynamics.
For instance, coupled kicked rotors (CKR) have been found to
display localization-delocalization behavior depending on the
coupling potential. For example, a CKR studied in
Ref.~\cite{DorFis1988} displays Anderson type of localization. A similar result
is also seen in CKR with different coupling potential
\cite{NotIemRosFazSilRus2018}. A CKR with a point interaction exhibits
dynamical localization of the center-of-mass momentum, which is destroyed for
the relative momentum \cite{QinAndParFla2017}.
In contrast, there are systems displaying the destruction of localization:
For example, a CKR with a certain coupling potential shows a diffusive growth
of the width of evolved state \cite{AdaTodIke1988}.
Similarly, for spatially confined pair of $\delta$-kicked rotors the
center-of-mass motion displays destruction of localization \cite{ParKim2003}. 
Moreover, for CKR either localization or diffusion
is reported, depending on the strength of coupling \cite{TolBal2009:p}.
Experimentally realized CKR shows a localization-delocalization 
transition~\cite{GadReeKriSch2013}.
Thus, the dynamics displayed by CKR depending on the coupling potential
is not yet fully understood and in particular the connection
to entropy production has not been elucidated.

In this paper we report on a surprising phenomenon in the
entanglement production of a pair of coupled kicked rotors on a cylinder,
which shows two distinct regimes
of entanglement growth, i.e.\ linear and logarithmic, as time progresses.
We show that this is tightly connected to a
localization-delocalization cross-over of time-evolved
states with an intermediate dynamical localization.
We also show that the logarithmic growth of the entanglement entropy
commences once the system displays normal diffusion at large times,
while before that a linear growth is found.
We further show that however weak the coupling is,
the rotor will eventually display normal diffusion at large times.
Analytically we calculate the growth of the linear entropy
which shows an initial linear behavior followed by a saturation.
The rate of linear growth is shown to depend quadratically
on the ratio of scaled Planck's constant
to the coupling strength rather than on the Lyapunov exponent.
Furthermore, we provide an analytical expression for the time
beyond which the logarithmic growth of the entanglement entropy starts.

A pair of coupled kicked rotors is a prototypical system
for studying the dynamics and entanglement between two particles.
Its Hamiltonian is given by
\begin{equation}
\begin{aligned}
H& = \frac{p_1^2}{2} + \frac{p_2^2}{2}
+\left[K_1~\cos(x_1) + K_2~\cos(x_2) \right.\\
&~~~ \left. + ~\xi_{12}~\cos(x_1-x_2)\right] \sum_n \delta(t-n)\\
& = H_1 + H_2 + H_{12},
\label{eq2}
\end{aligned}
\end{equation}
where $H_j =\frac{p_j^2}{2}+K_j~\cos(x_j) \sum_n \delta(t-n)$ represents the
Hamiltonian of each kicked rotor and
$H_{12} = \xi_{12}~\cos(x_1-x_2) \sum_n \delta(t-n)$ describes the coupling
between the two rotors.
Here $p_j$ is the momentum and $x_j$ the position of the $j$-th rotor.
The kicking strengths of the kick received by $j$-th kicked rotor is
$K_j$ and $\xi_{12}$ represents the coupling strength.
Considering the dynamics stroboscopically, i.e.\
at multiple integer times, one obtains a four-dimensional symplectic
map on a cylinder with periodic boundary conditions in the position coordinates.
Note that assuming in addition periodic boundary conditions
in the momentum coordinates one obtains the
four-dimensional coupled standard map~\cite{Fro1971,Fro1972,RicLanBaeKet2014}.

For $\xi_{12}=0$, the system represents two uncoupled kicked rotors.
If the kicking strengths $K_j$ of the individual rotors are
sufficiently large, their dynamics is chaotic
with a Lyapunov exponent of approximately $\ln (K_j/2)$~\cite{Chi1979}.
In the following numerical investigations we use $K_1=9.0$ and $K_2=10.0$
so that the classical dynamics is chaotic.
In this chaotic case,  the single or uncoupled kicked rotors
display normal diffusion, i.e.\ a linear growth of the mean energy
for an ensemble of initial conditions,
$\langle E \rangle = D_{\text{cl}}t$ \cite{CasChiIzrFor1979,Izr1990},
where $D_{\text{cl}}$ is the classical diffusion coefficient.

\begin{figure}[!h]
  \includegraphics[width=0.47\textwidth]{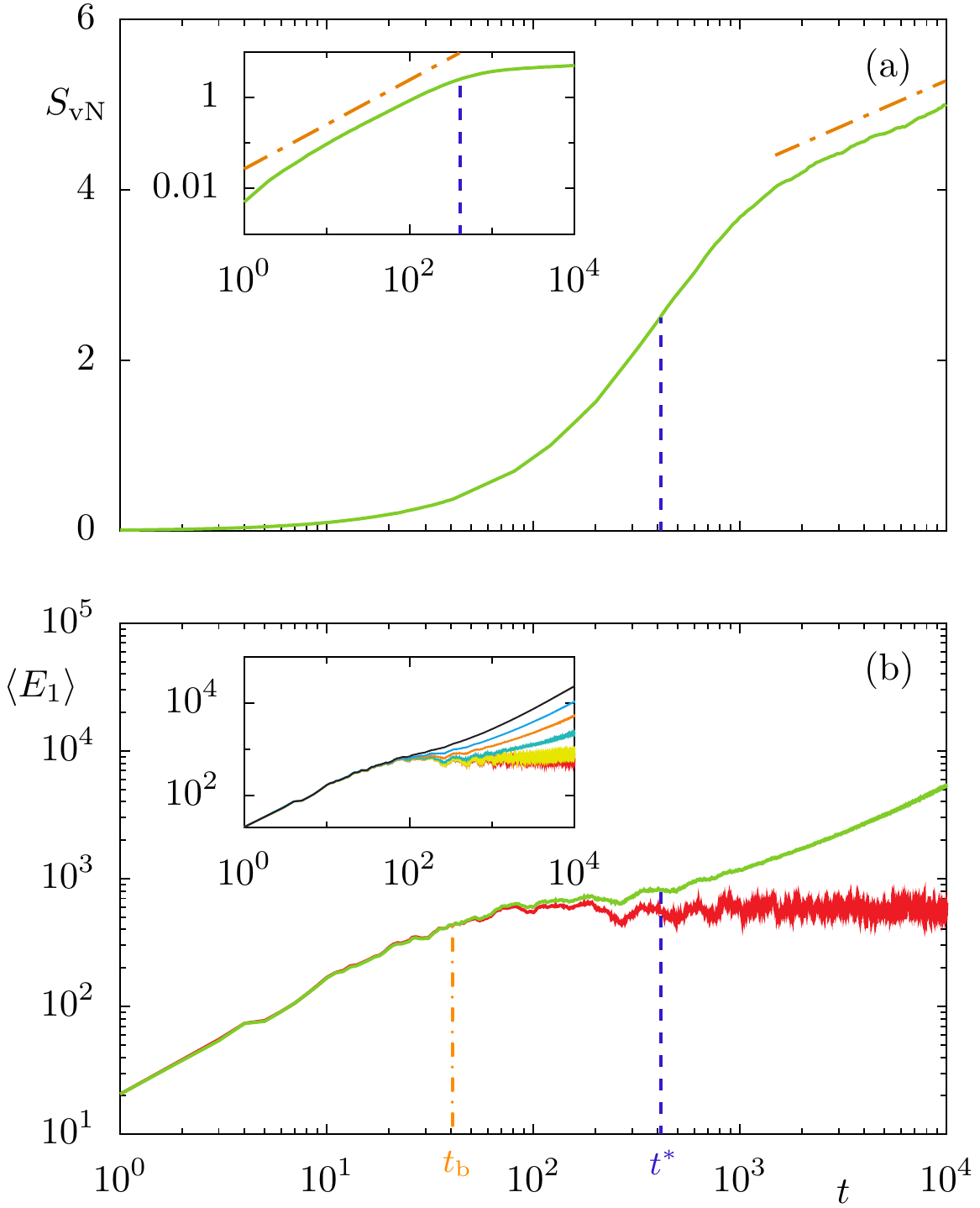}\vspace{0.3cm}
  \includegraphics[width=0.47\textwidth]{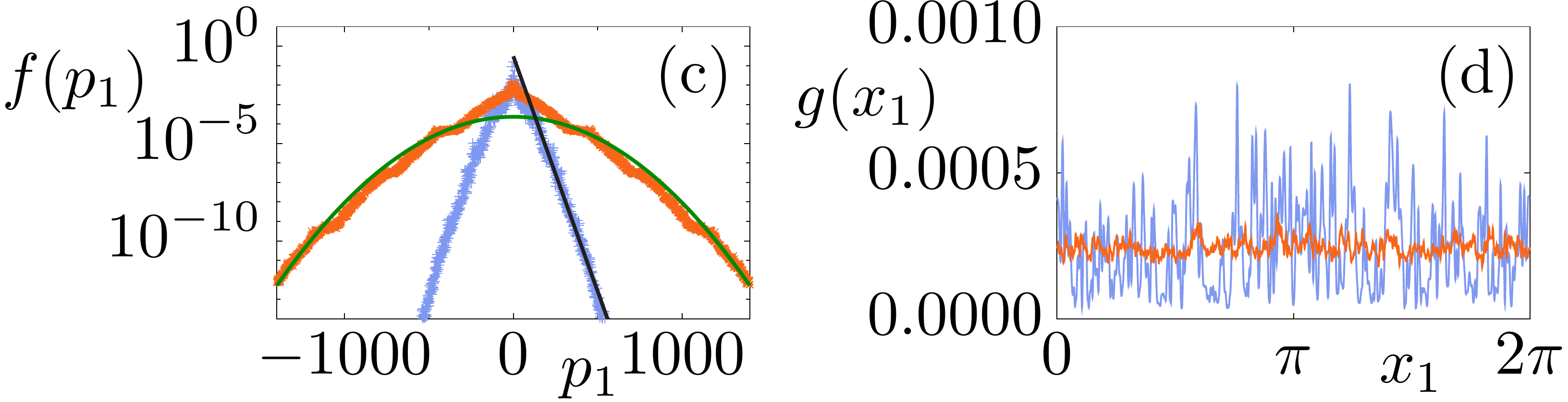}
  \caption{ (a)  Entanglement entropy $S_{\text{vN}}$ as function of time.
    The blue vertical dashed line represents the crossover time $t^*$
    at which the transition from linear to logarithmic occurs. Inset captures
    the linear regime. The orange
    dash dotted lines are
    linear fits to $S_{\text{vN}}$.
    (b) Mean energy growth
    $\langle E_1 \rangle$ of the first kicked rotor as a function of time.
    The orange dash dotted line indicates the break-time $\tb$.
    Inset shows
    $\langle E_1 \rangle$ for different coupling strengths,
    $\xi_{12}=0.0, 0.01, 0.03, 0.07, 0.1$ from bottom to top. (c) and (d)
    represent momentum and position distributions at two different times,
    $t=150$ (blue symbols) and $t=10000$ (orange symbols).
    The solid lines in (c) are Gaussian (green) and exponential (black) fits.
    All plots
    are for $K_1=9.0$, $K_2=10.0$, $\xi_{12}=0.05$ (for green solid line), and
    $\hbars=1.0$.
  }
\label{fig1}
\end{figure}

For the quantum dynamics of CKR,
the time evolution is given by the unitary operator
$U=(U_1\otimes U_2)U_{12}$, where
$U_j = \ue^{-\ui \frac{p_j^2}{2 \hbars}} \ue^{-\ui \frac{K_j}{\hbars}\cos(x_j)}$ and
$U_{12} = \ue^{-\ui \frac{\xi_{12}}{\hbars}\cos(x_1-x_2)}$,
such that $|\Psi(t)\rangle = U^t |\Psi(0)\rangle$ is
the time-evolved state at discrete time $t$
of the initial state $|\Psi(0)\rangle$.
In the following, we consider as initial state product states of the form
$|\Psi(0)\rangle = |\psi_1(0)\rangle \otimes |\psi_2(0)\rangle$,
where $|\psi_j(0)\rangle$ is a coherent state of the $j$-th rotor.
For the numerical calculations, the fast Fourier transform
is employed by expressing momentum and position values of each rotor
evaluated on discrete grids of the same size $N=2^l$ with $l=11$.
Note that in contrast to the normal diffusion displayed by a classical
single kicked rotor, its quantum dynamics shows a suppression of the
diffusion. This phenomenon is called dynamical localization, which is a phase
coherent effect analogous to Anderson localization observed in disordered
lattices \cite{GrePraFis1984}.
Also note, similar to the classical case, the pair of quantum kicked rotors
can also be considered with periodic boundary conditions
in momentum~\cite{ChaShi1986,Lak2001,RicLanBaeKet2014}
and the spectral properties and entanglement generation in this case
has been investigated in detail in
Refs.~\cite{SriTomLakKetBae2016,LakSriKetBaeTom2016,TomLakSriBae2018,
  PulLakSriBaeTom2020}.

The entanglement between the sub-systems given by
the two kicked rotors can be characterized by the von Neumann entropy
\begin{equation}
\begin{split}
S_{\text{vN}}(t)&=-\text{Tr}_1\left(\rho_1(t) \log \rho_1(t)\right),
\end{split}
\end{equation}
where $\rho_1(t) = \text{Tr}_2 (\rho(t))$ is the reduced density matrix obtained
by tracing out the contribution of second subsystem and
$\rho(t) = |\Psi(t)\rangle \langle \Psi(t)|$
is the total density matrix of the time-evolved state $|\Psi(t)\rangle$.

Figure \ref{fig1}(a) shows that the growth of $S_\text{vN}$
has two distinct regimes, linear and logarithmic.
Initially, $S_\text{vN}$ grows linearly
up to a cross-over time $t^*$, see the inset of Fig.~\ref{fig1}(a)
represented by a blue vertical dashed line.
We find that the rate of this linear growth depends on the coupling
strength following the relation $S_\text{vN} \sim \xi_{12}^{\beta}$. The value of
the exponent $\beta$ is numerically found to be approximately
$1.85$.  In this regime of linear growth, the rate turns out to be
independent of the kicking strengths when both rotors display classically
chaotic dynamics.
After the break-time $t^*$, the growth of $S_\text{vN}$ slows
down and shows a logarithmic dependence on time, i.e,
$S_\text{vN} = \tfrac{1}{2} \ln t + $ const.
The value of $S_\text{vN}$ at the onset of logarithmic growth, i.e.\
$S_\text{vN}(t^*)$,
depends on the quantum diffusion coefficient
which in turn depends
on the kicking strength. Additionally, it is found that
$S_\text{vN}(t^*)$ barely depends on the coupling strength (not shown).
Thus, the key finding is that the production of entanglement between the
rotors does not follow a single functional form and, remarkably,
in the two regimes a different dependency on the system parameters,
i.e.\ $\xi_{12}$ and $K_j$, is found.

To obtain a qualitative understanding of the observed entanglement
growth in terms of the underlying quantum dynamics,
let us consider the behavior of mean energy growth
$\langle E_1\rangle = \langle \Psi(t)| \frac{p_1^2}{2} |\Psi(t)\rangle$ and the
distributions $g(x_1) = |\langle \Psi|x_1\rangle|^2$ in position space
and $f(p_1) = |\langle \Psi|p_1\rangle|^2$ in momentum space
of the first rotor.
Figure~\ref{fig1}(b) shows that
there are three different regimes: Initially,
$\langle E_1\rangle$ grows linearly until the break time $\tb$,
which is indicated by a orange vertical dash dotted line in Fig.~\ref{fig1}(b). The
break-time is the time until which the quantum energy of a single kicked
rotor follows the classical energy growth \cite{Izr1990,MooRobBhaSunRai1995}.
Until this time the system builds
up its quantum correlations, which after $\tb$ leads to the emergence of an
intermediate dynamical localization (IDL)
for which $\langle E_1\rangle$ is essentially constant.
This IDL extends up to the cross-over time $t^*$,
indicated by a blue vertical dashed line in Fig.~\ref{fig1}(b).
Beyond $t^*$, the system displays normal diffusion,
$\langle E_1\rangle \sim t$. Moreover, it can be seen from the inset of
Fig.~\ref{fig1}(b), that the temporal extent of the IDL increases with
decreasing coupling $\xi_{12}$.
An important consequence of this
observation is that the system will always show normal diffusion at large times
for any non-vanishing coupling $\xi_{12}>0$.

Another significant observation is
that the normal diffusion
seen in Fig.~\ref{fig1}(b) is similar to classical diffusion.
This can be seen from Fig.~\ref{fig1}(c)
which shows that the momentum distribution
at time $t=10000$ is well described by a Gaussian.
Also the position distribution becomes very uniform with only
small quantum fluctuations, see Fig.~\ref{fig1}(d).
In contrast, the momentum distribution is exponentially localized in the IDL,
as illustrated at $t=150$ in Fig.~\ref{fig1}(c)
and the corresponding position distribution shows much larger fluctuations
as seen in Fig.~\ref{fig1}(d).

The Gaussian momentum and uniform position distributions are
typical features of a corresponding classical diffusive regime \cite{Izr1990}.
Thus, the appearance of normal diffusion suggests that the rotors provide noise to
each other which destroys quantum coherence.
Quantum coherence is the origin of the appearance of dynamical localization
for a single kicked rotor.
As a result, classical-like
behavior emerges which in turn gives rise to a slow growth of $S_\text{vN}$.
Thus, we can conclude that the linear regime of $S_\text{vN}$ appears
when the system has quantum correlations.
On the other hand, for $t>t^*$, where normal diffusion dominates,
complete loss of correlations gives rise to the logarithmic growth.

Now, we provide a theoretical explanation of the emergence of the
two regimes of $S_\text{vN}$ growth in the case of
weak coupling $\xi_{12}\ll 1$.
For this we consider the linear entropy
$S_{\text{lin}}(t) = \text{Tr}_1 \rho_1(t)^2$ which is analytically easier
tractable than the von Neumann entropy $S_\text{vN}$, but shows the same
characteristics.
To treat the initial time-dependence,
the key point is to consider that one rotor acts as
an environment to the other so that
we can rewrite the Hamiltonian in Eq.~\eqref{eq2} as $H = \HS +\HE + V(t)$,
where $\HS$ ($\HE$) represents the system (environment) Hamiltonian and $V(t)$
is the interaction.
The evolution of the total density matrix $\rho(t)$ in the
interaction picture is
\begin{equation}
\frac{\ud\rho(t)}{\ud t}=-\frac{\ui}{\hbars}\left[V(t),\rho(t)\right].
\end{equation}
As the initial state is a product state we have
$\rho(0)=\rhoS(0)\otimes \rhoE(0)$.
Performing formal integration and iteration and considering $\xi_{12} \ll 1$,
we arrive at
\begin{equation}
\begin{aligned}
\rho(t) =& \rho(0) - \frac{\text{i}\xi_{12}}{\hbars}\sum_{r=1}^t \left[\mathcal{F}(r),\rho(0)\right]\\
&+\left(\frac{\text{i}\xi_{12}}{\hbars}\right)^2\sum_{r=1}^t \sum_{s=1}^{r-1} \left[\mathcal{F}(s),\left[\mathcal{F}(r),\rho(0)\right]\right],
\end{aligned}
\label{rhosum}
\end{equation}
where $\mathcal{F}(r)=\cos(x_1(r)-x_2(r))$.
The summation instead of
integration that appears in Eq.~\eqref{rhosum} is due to the fact that coupling
acts only at integer times, i.e.\
$V(t) = \xi_{12}~\cos(x_1-x_2) \sum_n \delta(t-n)$.
The calculation of $\rho(t)$ is most conveniently done in position basis
as the interaction is in position space.
With $\rhoS(t)=\text{Tr}_E\rho(t)$ the computation of
$S_{\text{lin}}=1-\text{Tr}_{\text{S}}(\rhoS(t)^2)$ leads to
\begin{equation}
S_{\text{lin}}(t)=\left(\frac{\xi_{12}}{\hbars}\right)^2 C(t),
\label{slin_qn}
\end{equation}
where $C(t)=\sum_{r,s=1}^t C(r,s)$ and $C(r,s)$ represents the
correlation function at two different time steps. If both
rotors display classically chaotic dynamics, then $C(t)$ is independent of the
system parameters.
Furthermore, it is numerically found that
for small coupling, $C(t)$ depends linearly on time.
Thus Eq.~(\ref{slin_qn}) reveals that the
rate $\Gamma=\frac{\ud S_{\text{lin}}}{\ud t}$ depends only on the ratio
$\frac{\xi_{12}}{\hbars}$ rather than on the kicking strengths $K_j$.
This implies that for weak coupling the initial temporal growth
of $S_{\text{lin}}$ does not depend on the strength of chaos
of the individual rotors.

To determine the behavior of the linear entropy at large times,
i.e.\ for $t> t^*$, we employ that the time-evolved initial state
becomes on average Gaussian in momentum and uniform in position space,
see Fig.~\ref{fig1}(c) and (d).
Using the Husimi function $\mathcal{H}(x_1,p_1)$ one
can express the linear entropy as \cite{NagLahGho2001}
\begin{equation}
S_{\text{lin}}=1-\int \mathcal{H}(x_1,p_1)^2 \frac{dp_1 dx_1}{2\pi \hbars}.
\label{slin}
\end{equation}
We approximate the Husimi distribution of the time-evolved state by
$\mathcal{H}(x_1,p_1)=\frac{\hbars}{\sqrt{2\pi \Dq t}}
\exp\left(-\frac{p_1^2}{2 \Dq t}\right)$, where
$\Dq$ is the quantum diffusion coefficient.
Inserting $\mathcal{H}(x_1,p_1)$ in Eq.~\eqref{slin} gives
\begin{equation}
S_{\text{lin}}=1-\frac{\hbars}{\sqrt{4\pi \Dq}}t^{-1/2}.
\label{slin_cl}
\end{equation}
Equation~\eqref{slin_cl} reveals that $S_{\text{lin}}$ in the regime of
linear growth of $S_\text{vN}$ depends on $\Dq$ and
shows that $S_{\text{lin}}$ saturates at large times.

\begin{figure}[!h]
  \includegraphics[width=0.47\textwidth]{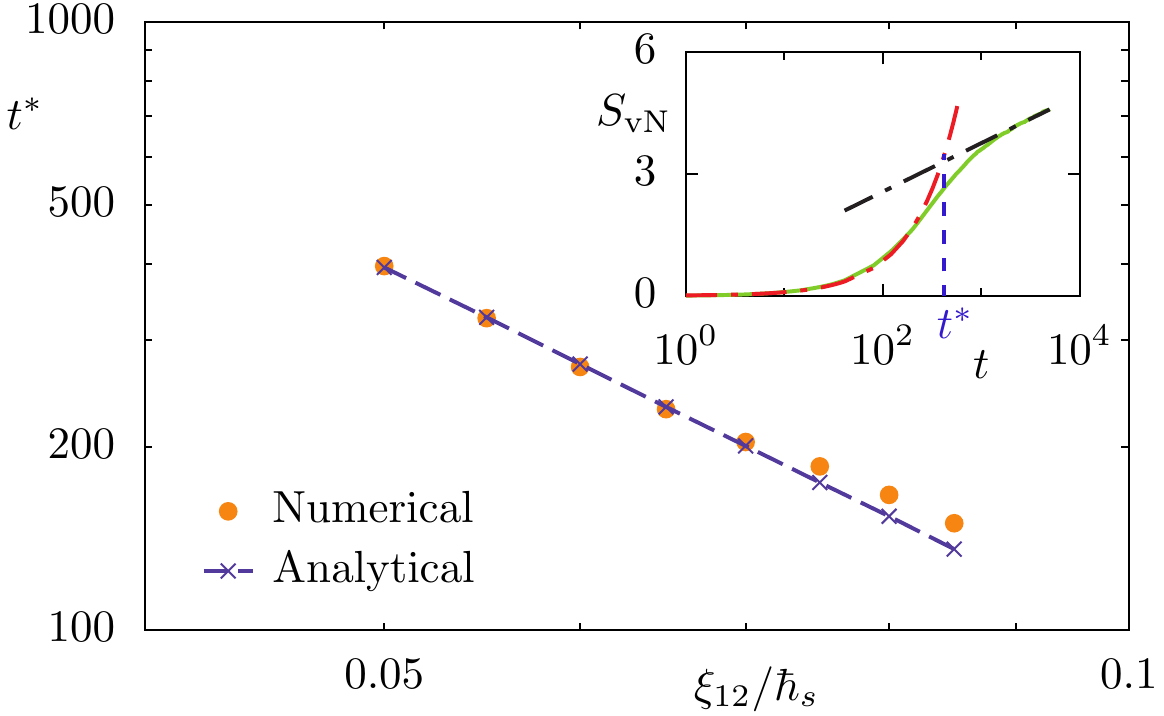}
  \caption{Dependence of the cross-over time $t^*$ on the ratio
    of the coupling strength $\xi_{12}$ to the
    scaled Planck's constant $\hbars$.
    The analytical result (blue dashed line with crosses)
    is compared to the numerical results (orange circles).
    The inset sketches the
    procedure to calculate $t^*$ numerically. Parameters are
    $K_1=9.0$, $K_2=10.0$, and $\hbars=1.0$.}
\label{fig3}
\end{figure}

Equating the two expressions for $S_{\text{lin}}$ obtained in Eq.~\eqref{slin_qn}
and Eq.~\eqref{slin_cl} at $t=t^*$, at which the cross-over occurs,
and solving for $t^*$, we obtain
\begin{equation}
  t^* = \frac{\hbars^2}{3 \xi_{12}^2}
        \left[2+\frac{1}{\mathcal{G}(\xi_{12},\Dq)}
              +\mathcal{G}(\xi_{12},\Dq)\right],
\label{crosst}
\end{equation}
where
\begin{equation*}
  \mathcal{G}(\xi_{12},\Dq) =
   \left[-1+\frac{27}{8\pi}\frac{\xi_{12}^2}{\Dq}
         +\frac{3^{\frac{3}{2}}\xi_{12}}{2}\sqrt{\frac{-1}{\pi \Dq}
         +\frac{27}{16 \pi^2}\frac{\xi_{12}^2}{\Dq^2}}\right]^{\frac{1}{3}}.
\end{equation*}
Hence, $t^*$ depends on the coupling strength $\xi_{12}$, the scaled Planck's
constant $\hbars$, and the quantum diffusion coefficient $\Dq$.
Equation~\eqref{crosst} shows that with decreasing
$\hbars$ the value of $t^*$ decreases. This is because quantum correlations
vanish in the semi-classical limit. Also, with increasing $\xi_{12}$,
the cross-over time $t^*$ decreases.
This relates to the fact that the coupling between the subsystems
destroys the coherence in the system.
Figure~\ref{fig3} compares the analytical result~\eqref{crosst}
with numerical result and a very good agreement is found.
This shows that, however small
the coupling is, the system will eventually show normal diffusion.
It is also interesting to note that for
a noisy kicked rotor with noise strength $\epsilon$, the diffusion commences at a time-scale
$\frac{\hbars^2}{\epsilon^2}$~\cite{OttAntHan1984}.

\begin{figure}[!h]
  \includegraphics[width=0.47\textwidth]{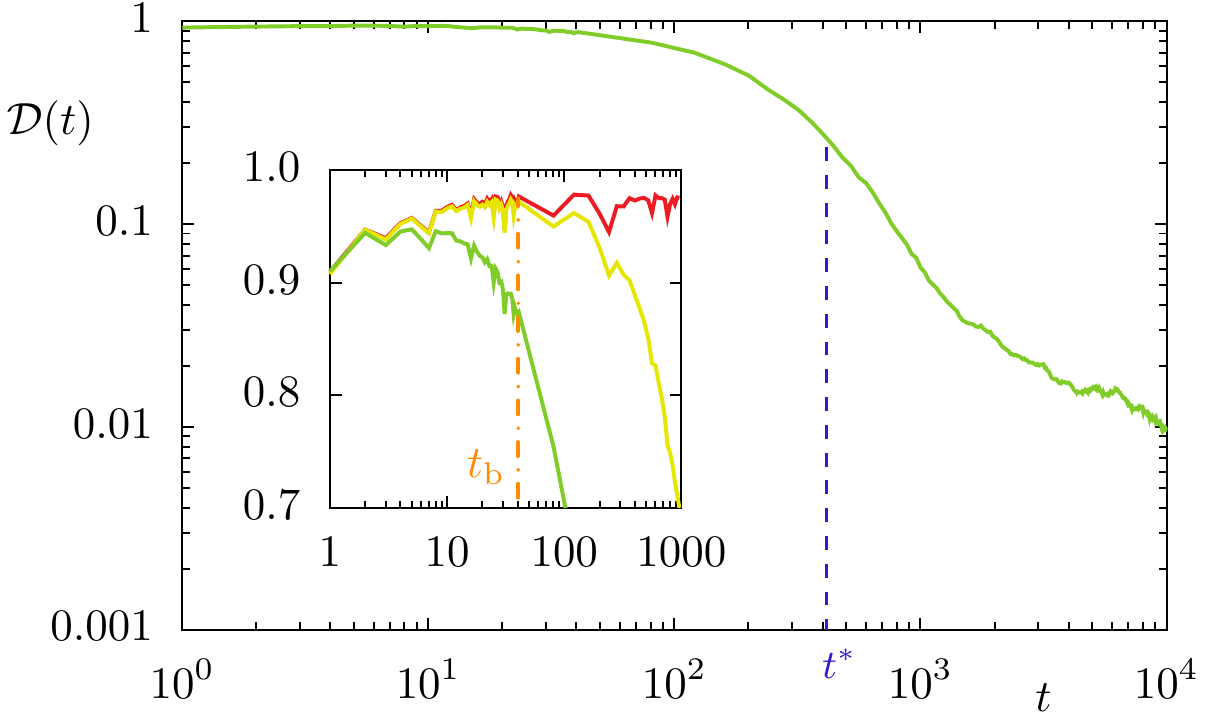}
  \caption{Decay of coherence of the first kicked rotor for
    $\xi_{12}=0.05$. The blue vertical dashed line represents $t^*$.
    The red, yellow, and green
    curves in the inset show the decoherence for
    $\xi_{12}=0.0,~0.01,~0.05$, respectively.
    The orange vertical dash dotted line indicates the break time $\tb$.
    Parameters are $K_1=9.0$, $K_2=10.0$, and $\hbars=1.0$.}
	\label{fig4}
\end{figure}

As discussed, coherence plays a central role in the emergence
of the different regimes of the entanglement growth as characterized
by $S_\text{vN}$ or $S_{\text{lin}}$.
To investigate the decay of coherence and examine the nature of
noise provided by one rotor to the other, we study
the decay of the off-diagonal elements of $\rho_1$.
The off-diagonal elements represent the interference between the system and the
environment
and their decay indicates the loss of coherence \cite{Zur2003}.
We quantify this decoherence by calculating
$\mathcal{D}(t)= \sum_{i\neq j} \rho_1^{ij}(t)$,
which is shown in Fig.~\ref{fig4}.
Initially $\mathcal{D}(t)$ is close to one and follows
an exponential decay until $t<t^*$. 
The exponential decay of coherence is also observed 
in a kicked rotor system
with random noise \cite{WhiRudHoo2014,PauSarVisManSanRap2019}.
This suggests that one rotor
provides random noise to the other and thus effectively acts as an environment.
Now, around $t=t^*$  in Fig.~\ref{fig4}, one observes an extended transition
and finally a power-law decay.
The exponent of the power-law is numerically found to be
approximately $0.5$.
This slow decay of coherence
implies that the quantum system enters the classical-like regime
as illustrated in Fig.~\ref{fig1}(b) and will require an arbitrary large time
to actually behave like a classical system.

A closer look at the initial time-dependence of the decoherence,
as shown in the inset of
Fig.~\ref{fig4}, reveals an initial production of coherence
for uncoupled rotors (red curve) until the break-time $\tb$
and then saturation to a constant value.
This initial increase can also be observed
in a weak coupling situation (yellow curve)
for which $\tb\ll t^*$ and because of that, a signature of IDL is observed as
in Fig.~\ref{fig1}(b) for $t>\tb$.
Even for coupling $\xi_{12}=0.05$,
an initial increase in coherence for a small time interval $t<\tb$
can be observed (green curve in the inset of Fig.~\ref{fig4})
which corresponds to the appearance of a short IDL
in Fig.~\ref{fig1}(b).
However, for strong coupling, the cross-over time $t^*$ becomes so small
that an initial production of coherence is not possible.
Thus, the coherence decays from the very beginning.

To summarize, for a pair of coupled kicked rotors
we demonstrate that the entanglement entropy
shows two distinct regimes, initially linear growth followed by
a logarithmic increase.
The logarithmic regime sets in when the time-evolved state
shows a Gaussian profile in momentum space and is uniform on average
in position space.
This regime can be considered as a kind of classical behavior
caused by one rotor acting as a noisy environment to the other.
This leads to an exponential decoherence, which is clearly confirmed
by the numerical results.
The cross-over time $t^*$ between linear and logarithmic
behavior of the von Neumann entropy is computed
using the linear entropy. Explicit expressions for $S_{\text{lin}}$
in  both regimes are obtained and excellent agreement of the prediction
for $t^*$ with numerics is found.
Thus, we show that entanglement entropy allows to distinguish
two completely different dynamics, quantum and classical-like.
It would be very interesting to experimentally investigate this
CKR, for example using ultra-cold atoms.

\acknowledgments

We thank Roland Ketzmerick, Arul Lakshminarayan, and David Luitz
for useful discussions.



\begin{thebibliography}{10}
	\newcommand{\enquote}[1]{``#1''}
	\providecommand{\url}[1]{\texttt{#1}}
	\providecommand{\urlprefix}{URL }
	\providecommand{\eprint}[2][]{\url{#2}}
	
	\bibitem{HorHorHorHor2009b}
	R.~Horodecki, P.~Horodecki, M.~Horodecki, and K.~Horodecki, \emph{Quantum
		entanglement}, Rev.~Mod.~Phys. \textbf{81}, 865 (2009).
	
	\bibitem{Laf2016}
	N.~Laflorencie, \emph{Quantum entanglement in condensed matter systems},
	Phys.~Rep. \textbf{646}, 1 (2016).
	
	\bibitem{YanChaHamMuc2015}
	Z.-C. Yang, C.~Chamon, A.~Hamma, and E.~R. Mucciolo, \emph{Two-component
		structure in the entanglement spectrum of highly excited states},
	Phys.~Rev.~Lett. \textbf{115}, 267206 (2015).
	
	\bibitem{BarPolMoo2012}
	J.~H. Bardarson, F.~Pollmann, and J.~E. Moore, \emph{Unbounded growth of
		entanglement in models of many-body localization}, Phys.~Rev.~Lett.
	\textbf{109}, 017202 (2012).
	
	\bibitem{SerPapAba2013}
	M.~Serbyn, Z.~Papi{\'c}, and D.~A. Abanin, \emph{Universal slow growth of
		entanglement in interacting strongly disordered systems}, Phys.~Rev.~Lett.
	\textbf{110}, 260601 (2013).
	
	\bibitem{VosAlt2013}
	R.~Vosk and E.~Altman, \emph{Many-body localization in one dimension as a
		dynamical renormalization group fixed point}, Phys.~Rev.~Lett. \textbf{110},
	067204 (2013).
	
	\bibitem{ZniProPre2008}
	M.~\v{Z}nidari\v{c}, T.~Prosen, and P.~Prelov\v{s}ek, \emph{Many-body
		localization in the {Heisenberg} {$XXZ$} magnet in a random field},
	Phys.~Rev.~B \textbf{77}, 064426 (2008).
	
	\bibitem{LukRisSchTaiKauChoKheLeoGre2019}
	A.~Lukin, M.~Rispoli, R.~Schittko, M.~E. Tai, A.~M. Kaufman, S.~Choi,
	V.~Khemani, J.~L{\'e}onard, and M.~Greiner, \emph{Probing entanglement in a
		many-body -- localized system}, Science \textbf{364}, 256 (2019).
	
	\bibitem{LuiLafAle2016}
	D.~J. Luitz, N.~Laflorencie, and F.~Alet, \emph{Extended slow dynamical regime
		close to the many-body localization transition}, Phys.~Rev.~B \textbf{93},
	060201 (2016).
	
	\bibitem{SinBarPol2016}
	R.~Singh, J.~H. Bardarson, and F.~Pollmann, \emph{Signatures of the many-body
		localization transition in the dynamics of entanglement and bipartite
		fluctuations}, New J. Phys. \textbf{18}, 023046 (2016).
	
	\bibitem{SchLanRooDal2013}
	J.~Schachenmayer, B.~P. Lanyon, C.~F. Roos, and A.~J. Daley, \emph{Entanglement
		growth in quench dynamics with variable range interactions}, Phys.~Rev.~X
	\textbf{3}, 031015 (2013).
	
	\bibitem{LerPap2020}
	A.~Lerose and S.~Pappalardi, \emph{Origin of the slow growth of entanglement
		entropy in long-range interacting spin systems}, Physical Review Research
	\textbf{2}, 012041 (2020).
	
	\bibitem{LerMarGamSil2019}
	A.~Lerose, J.~Marino, A.~Gambassi, and A.~Silva, \emph{Prethermal quantum
		many-body kapitza phases of periodically driven spin systems}, Phys.~Rev.~B
	\textbf{100}, 104306 (2019).
	
	\bibitem{vanLevGar2015}
	M.~{van Horssen}, E.~Levi, and J.~P. Garrahan, \emph{Dynamics of many-body
		localization in a translation-invariant quantum glass model}, Phys.~Rev.~B
	\textbf{92}, 100305 (2015).
	
	\bibitem{NahRuhVijHaa2017}
	A.~Nahum, J.~Ruhman, S.~Vijay, and J.~Haah, \emph{Quantum entanglement growth
		under random unitary dynamics}, Phys.~Rev.~X \textbf{7}, 031016 (2017).
	
	\bibitem{CalCar2005}
	P.~Calabrese and J.~Cardy, \emph{Evolution of entanglement entropy in
		one-dimensional systems}, J.~Stat.~Mech. \textbf{2005}, P04010 (2005).
	
	\bibitem{BiaHacYok2018}
	E.~Bianchi, L.~Hackl, and N.~Yokomizo, \emph{Linear growth of the entanglement
		entropy and the {Kolmogorov}-{Sinai} rate}, J.~High Energy Phys. \textbf{03},
	025 (2018).
	
	\bibitem{ZurPaz1994}
	W.~H. Zurek and J.~P. Paz, \emph{Decoherence, chaos, and the second law},
	Phys.~Rev.~Lett. \textbf{72}, 2508 (1994).
	
	\bibitem{MilSar1999}
	P.~A. Miller and S.~Sarkar, \emph{Entropy production, dynamical localization
		and criteria for quantum chaos in the open quantum kicked rotor},
	Nonlinearity \textbf{12}, 419 (1999).
	
	\bibitem{FujTanMiy2003}
	H.~Fujisaki, A.~Tanaka, and T.~Miyadera, \emph{Dynamical aspects of quantum
		entanglement for coupled mapping systems}, J.~Phys.~Soc.~Jpn. \textbf{72},
	111 (2003).
	
	\bibitem{FujMiyTan2003}
	H.~Fujisaki, T.~Miyadera, and A.~Tanaka, \emph{Dynamical aspects of quantum
		entanglement for weakly coupled kicked tops}, Phys.~Rev.~E \textbf{67},
	066201 (2003).
	
	\bibitem{DorFis1988}
	E.~Doron and S.~Fishman, \emph{Anderson localization for a two-dimensional
		rotor}, Phys.~Rev.~Lett. \textbf{60}, 867 (1988).
	
	\bibitem{NotIemRosFazSilRus2018}
	S.~Notarnicola, F.~Iemini, D.~Rossini, R.~Fazio, A.~Silva, and A.~Russomanno,
	\emph{From localization to anomalous diffusion in the dynamics of coupled
		kicked rotors}, Phys.~Rev.~E \textbf{97}, 022202 (2018).
	
	\bibitem{QinAndParFla2017}
	P.~Qin, A.~Andreanov, H.~C. Park, and S.~Flach, \emph{Interacting ultracold
		atomic kicked rotors: Loss of dynamical localization}, Sci.~Rep. \textbf{7},
	41139 (2017).
	
	\bibitem{AdaTodIke1988}
	S.~Adachi, M.~Toda, and K.~Ikeda, \emph{Quantum-classical correspondence in
		many-dimensional quantum chaos}, Phys.~Rev.~Lett. \textbf{61}, 659 (1988).
	
	\bibitem{ParKim2003}
	H.-K. Park and S.~W. Kim, \emph{Decoherence from chaotic internal dynamics in
		two coupled $\delta$-function-kicked rotors}, Phys.~Rev.~A \textbf{67},
	060102 (2003).
	
	\bibitem{TolBal2009:p}
	B.~Toloui and L.~E. Ballentine, \emph{Quantum localization for two coupled
		kicked rotors}, {arXiv}:0903.4632 {[}quant-ph{]}  (2009).
	
	\bibitem{GadReeKriSch2013}
	B.~Gadway, J.~Reeves, L.~Krinner, and D.~Schneble, \emph{Evidence for a
		quantum-to-classical transition in a pair of coupled quantum rotors}, Phys.
	Rev. Lett. \textbf{110}, 190401 (2013).
	
	\bibitem{Fro1971}
	C.~{Froeschle}, \emph{On the number of isolating integrals in systems with
		three degrees of freedom}, Astrophys.~Space Sci. \textbf{14}, 110 (1971).
	
	\bibitem{Fro1972}
	C.~{Froeschl\'e}, \emph{Numerical study of a four-dimensional mapping},
	Astron.~\&~Astrophys. \textbf{16}, 172 (1972).
	
	\bibitem{RicLanBaeKet2014}
	M.~Richter, S.~Lange, A.~B\"acker, and R.~Ketzmerick, \emph{Visualization and
		comparison of classical structures and quantum states of four-dimensional
		maps}, Phys.~Rev.~E \textbf{89}, 022902 (2014).
	
	\bibitem{Chi1979}
	B.~V. {Chirikov}, \emph{{A universal instability of many-dimensional oscillator
			systems}}, Phys.~Rep. \textbf{52}, 263 (1979).
	
	\bibitem{CasChiIzrFor1979}
	G.~Casati, B.~Chirikov, F.~Izraelev, and J.~Ford, \emph{Stochastic behavior of
		a quantum pendulum under a periodic perturbation}, in G.~Casati and J.~Ford
	(editors) \enquote{Stochastic Behavior in Classical and Quantum Hamiltonian
		Systems}, volume~93 of \emph{Lect.~Notes Phys.}, 334, Springer Berlin /
	Heidelberg, Berlin (1979).
	
	\bibitem{Izr1990}
	F.~M. Izrailev, \emph{Simple models of quantum chaos: Spectrum and
		eigenfunctions}, Phys.~Rep. \textbf{196}, 299 (1990).
	
	\bibitem{GrePraFis1984}
	D.~R. Grempel, R.~E. Prange, and S.~Fishman, \emph{Quantum dynamics of a
		nonintegrable system}, Phys. Rev. A \textbf{29}, 1639 (1984).
	
	\bibitem{ChaShi1986}
	S.-J. Chang and K.-J. Shi, \emph{Evolution and exact eigenstates of a resonant
		quantum system}, Phys.~Rev.~A \textbf{34}, 7 (1986).
	
	\bibitem{Lak2001}
	A.~Lakshminarayan, \emph{Entangling power of quantized chaotic systems},
	Phys.~Rev.~E \textbf{64}, 036207 (2001).
	
	\bibitem{SriTomLakKetBae2016}
	S.~C.~L. Srivastava, S.~Tomsovic, A.~Lakshminarayan, R.~Ketzmerick, and
	A.~B\"acker, \emph{Universal scaling of spectral fluctuation transitions for
		interacting chaotic systems}, Phys.~Rev.~Lett. \textbf{116}, 054101 (2016).
	
	\bibitem{LakSriKetBaeTom2016}
	A.~Lakshminarayan, S.~C.~L. Srivastava, R.~Ketzmerick, A.~B{\"a}cker, and
	S.~Tomsovic, \emph{Entanglement and localization transitions in eigenstates
		of interacting chaotic systems}, Phys.~Rev.~E \textbf{94}, 010205(R) (2016).
	
	\bibitem{TomLakSriBae2018}
	S.~Tomsovic, A.~Lakshminarayan, S.~C.~L. Srivastava, and A.~B\"acker,
	\emph{Eigenstate entanglement between quantum chaotic subsystems: Universal
		transitions and power laws in the entanglement spectrum}, Phys.~Rev.~E
	\textbf{98}, 032209 (2018).
	
	\bibitem{PulLakSriBaeTom2020}
	J.~J. Pulikkottil, A.~Lakshminarayan, S.~C.~L. Srivastava, A.~B{\"a}cker, and
	S.~Tomsovic, \emph{Entanglement production by interaction quenches of quantum
		chaotic subsystems}, Phys.~Rev.~E \textbf{101}, 032212 (2020).
	
	\bibitem{MooRobBhaSunRai1995}
	F.~L. Moore, J.~C. Robinson, C.~F. Bharucha, B.~Sundaram, and M.~G. Raizen,
	\emph{Atom optics realization of the quantum $\delta$-kicked rotor},
	Phys.~Rev.~Lett. \textbf{75}, 4598 (1995).
	
	\bibitem{NagLahGho2001}
	S.~Nag, A.~Lahiri, and G.~Ghosh, \emph{Entropy production due to coupling to a
		heat bath in the kicked rotor problem}, Phys.~Lett.~A \textbf{292}, 43
	(2001).
	
	\bibitem{OttAntHan1984}
	E.~Ott, T.~M. Antonsen, and J.~D. Hanson, \emph{Effect of noise on
		time-dependent quantum chaos}, Phys.~Rev.~Lett. \textbf{53}, 2187 (1984).
	
	\bibitem{Zur2003}
	W.~H. Zurek, \emph{Decoherence, einselection, and the quantum origins of the
		classical}, Rev.~Mod.~Phys. \textbf{75}, 715 (2003).
	
	\bibitem{WhiRudHoo2014}
	D.~H. White, S.~K. Ruddell, and M.~D. Hoogerland, \emph{Phase noise in the
		delta kicked rotor: From quantum to classical}, New J. Phys. \textbf{16},
	113039 (2014).
	
	\bibitem{PauSarVisManSanRap2019}
	S.~Paul, S.~Sarkar, C.~Vishwakarma, J.~Mangaonkar, M.~S. Santhanam, and
	U.~Rapol, \emph{Nonmonotonic diffusion rates in an atom-optics {L\'evy}
		kicked rotor}, Phys.~Rev.~E \textbf{100}, 060201 (2019).
	
\end{thebibliography}
\end{document}